\documentstyle[twocolumn,prl,aps]{revtex}
\draft
\begin{document}
\twocolumn[
\title{Bethe Ansatz Approach to Thermodynamics of Superconducting
Magnetic Alloys}
\author{Valery I. Rupasov\cite{pa}}
\address{Department of Physics, University of Toronto, Toronto,
Ontario, Canada M5S 1A7}
\date{\today}
\maketitle
\widetext\leftskip=1.5cm\rightskip=1.5cm\nointerlineskip\small
\hspace{2.5mm}
We derive thermodynamic Bethe ansatz equations for a model
describing an $U\to\infty$ Anderson impurity embedded in a
BCS superconductor. The equations are solved analytically
in the zero-temperature limit, $T=0$. It is shown that the
impurities depress superconductivity in the Kondo limit,
however at $T=0$ the system remains in the superconducting
state for any impurity concentration. In the mixed-valence
regime, an impurity contribution to the density of states
of the system near the Fermi level overcompensates a Cooper
pairs weakening, and superconductivity is enhanced.

]
\narrowtext

Since the pioneering work of Abrikosov and Gor'kov (AG)
\cite{AG}, the problem of superconducting magnetic alloys
has been the subject of many early \cite{SMA} and more recent
\cite{NSMA} studies. Almost all of the theoretical methods
developed to attack the Kondo problem in normal metals, from
perturbative approaches to Wilson's numerical renormalization
group, have been generalized to the case of superconductors.
Perhaps the only exception is the Bethe ansatz (BA) technique,
which solves the Kondo problem in normal metals exactly, but
cannot be straightforwardly generalized to the case of dilute
superconducting alloys. The basic theoretical models describing
magnetic impurities in normal metals, such as the s-d (Kondo)
and Anderson models, are integrable under {\em two additional
conditions}: (i) an electron-impurity coupling is assumed to be
energy independent, and (ii) a band electron dispersion, $E_k$,
can be linearized around the Fermi level, $E_k\simeq v_F(k-k_F)$,
where $k_F$ and $v_F$ are the Fermi momentum and velocity,
respectively \cite{BA}. Since a carrier dispersion in the
superconducting state cannot be linearized near the Fermi
level, these conditions eliminate superconductivity from
the BA analysis of the behavior of magnetic alloys.

However, it has recently been discerned \cite{RS} that
the basic ``impurity'' models of quantum optics, describing
a system of Bose particles with a nonlinear dispersion coupled
to two-level atoms, exhibit {\em hidden integrability} and are
thus exactly diagonalized by BA. One of the most exciting
potential applications of the approach developed may be an
extension of the BA method to the Kondo problem in
superconductors and other Fermi systems (e. g., gapless Fermi
systems \cite{GFS}) with an essentially nonlinear dispersion
of charge carriers.

In this Letter, we employ hidden integrability of a model
describing an Anderson impurity with an infinitely large
Coulomb repulsion on an impurity orbital embedded in a BCS
superconductor \cite{R} to study the thermodynamic properties
of the system. In the standard manner \cite{BA}, we find
a set of basic thermodynamic equations for the energies of
elementary excitations of the system. In terms of these
equations, we derive an exact equation for the order parameter
of the superconducting phase transition \cite{N}, $\Delta$,
minimizing the thermodynamic potential of the system, $\Omega$,
with respect to $\Delta$, $\delta\Omega/\delta\Delta=0$. While
at finite temperatures the basic equations require a numerical
analysis, in the zero-temperature limit, $T=0$, they are solved
analytically, giving an exact expression for the impurity
contribution to the parameter $\Delta$.

At $T=0$, the order parameter is given by the expression
$\Delta=\Delta_0\exp{(-\mu_{\rm imp})}$, where $\Delta_0$
is the order parameter in the absence of impurities and
the parameter $\mu_{\rm imp}$ [see Eq. (15)] describes the
impurity contribution. The magnitude and sign of $\mu_{\rm imp}$
are determined completely by the position of the impurity energy
level, $\epsilon_d$, with respect to the Fermi energy of the
host metal, $\epsilon_F$. In the Kondo limit, where $\epsilon_d$
lies much below $\epsilon_F$, $\mu_{\rm imp}$ is positive, and
hence magnetic impurities depress superconductivity. However, the
system remains in the superconducting state at any concentration
of impurities. In the mixed-valence regime, where $\epsilon_d$
lies near the Fermi level, the parameter $\mu_{\rm imp}$
becomes negative. An impurity contribution to the density
of states of the system near the Fermi level dominates over
a Cooper pairs weakening, and the Anderson impurities enhance
superconductivity.

Throughout the paper, the BA technique and many results of the
exact solution of the Anderson model in normal metals are often
used with no special references. All needed details can be found
in excellent comprehensive reviews \cite{BA} and references therein.

We start with a Hamiltonian that includes the Hamiltonians of
the BCS and Anderson models,
\begin{eqnarray}
H&=&\sum_{k,\sigma}E_k a^\dagger_{k\sigma}a_{k\sigma}-
\sum_{k}\left(\Delta a^\dagger_{k\uparrow}a^\dagger_{k\downarrow}+
\Delta^* a_{k\downarrow}a_{k\uparrow}\right)\nonumber\\
&&+\Delta^2/{\rm g}+\sum_{k,\sigma}v_k
\left(a^\dagger_{k\sigma}d_\sigma+d^\dagger_\sigma a_{k\sigma}\right)
\nonumber\\
&&+\epsilon_d\sum_{\sigma}d^\dagger_\sigma d_\sigma
+Ud^\dagger_\uparrow d_\uparrow d^\dagger_\downarrow d_\downarrow,
\end{eqnarray}
where we have used the standard spherical harmonic representation
for band electron operators. The Fermi operator $a^\dagger_{k\sigma}$
creates a conduction electron with the momentum modulus $k$,
spin $\sigma=\uparrow,\downarrow$ and the energy
$E_k=\epsilon_k-\epsilon_F$, where $\epsilon_k$ is the kinetic
energy. Only the $s$-wave is assumed to be coupled to the impurity,
therefore all other partial waves have been dropped. An electron
localized in the impurity orbital is described by the Fermi
operators $d_\sigma$. The fourth term of Eq. (1) represents
the hybridization of the band and impurity level electrons
with the matrix element $v_k$, while the Coulomb repulsion on
the impurity orbital is described by the last term. The parameter
$\Delta={\rm g}\sum_{k}\langle a_{k\downarrow}a_{k\uparrow}\rangle$
is assumed to result from the Cooper pairing phenomenon with
positive coupling constant ${\rm g}$.

Diagonalization of the BCS part of Eq. (1) by the Bogoliubov-Valatin
unitary transform \cite{T} gives
\begin{eqnarray}
H&=&E_{\rm BCS}+\sum_{k\sigma}\omega_k c^\dagger_{k\sigma}c_{k\sigma}
+v\sum_{k\sigma}\left(d^\dagger_\sigma c_{k\sigma}
+c^\dagger_{k\sigma}d_\sigma\right)\nonumber\\
&&+\epsilon_d\sum_{\sigma}d^\dagger_\sigma d_\sigma
+Ud^\dagger_\uparrow d_\uparrow d^\dagger_\downarrow d_\downarrow,
\end{eqnarray}
where $\omega_k=-\sqrt{k^2+\Delta^2}$, for $k<0$,
$\omega_k=\sqrt{k^2+\Delta^2}$ for $k>0$, and
$
E_{\rm BCS}=\sum_{k}(k-\omega_k)+\Delta^2/{\rm g}.
$
For simplicity, we have linearized the band spectrum of the host
metal in the normal state around the Fermi level and set $v_F=1$.
The electron momentum and energy are taken relative to the Fermi
values. We have also set $k=k_F$ in both the hybridization matrix
element, $v=v(k_F)$, and in the coefficients of the unitary transform.

Moreover, to apply the BA method to the Hamiltonian (2), we
have omitted the terms $d^\dagger_\sigma c^\dagger_{k\sigma}$ and
$d_\sigma c_{k\sigma}$, which do not conserve the number of
excitations in the system; these terms are assumed to lead only
to insignificant corrections. The bare vacuum of the model
is then defined by $c_{k\sigma}|0\rangle=d_\sigma|0\rangle=0$.
To obtain the ground state of the system in the absence of the
impurity, one thus needs to fill all states with $k<0$. In the
normal state, $\Delta=0$, Eq. (2) reduces to the integrable
version of the Anderson model diagonalized by Wiegmann \cite{W}.

In what follows, we confine ourselves to very large values
of $U$, $\epsilon_d+U>{\cal D}^{(+)}$, where ${\cal D}^{(+)}$
is the upper edge of the band, so that double occupancy of
the impurity level is excluded. The eigenvalues of the model
Hamiltonian (2) are then found from the following Bethe ansatz
equations (BAE) \cite{R}:
\begin{mathletters}
\begin{eqnarray}
&&\exp{(ik_jL)}\,
\frac{h_j-\epsilon_d/2\Gamma-i/2}{h_j-\epsilon_d/2\Gamma+i/2}=
\prod_{\alpha=1}^{M}
\frac{h_j-\lambda_\alpha-i/2}{h_j-\lambda_\alpha+i/2}\\
&&\prod_{j=1}^{N}\frac{\lambda_\alpha-h_j-i/2}{\lambda_\alpha-h_j+i/2}=
-\prod_{\beta=1}^{M}\frac{\lambda_\alpha-\lambda_\beta-i}
{\lambda_\alpha-\lambda_\beta+i},
\end{eqnarray}
where $E=\sum_{j}\omega(k_j)$ is the eigenenergy, $N$ is the
total number of particles in the interval $L$, and $M$ is the
number of particles with spin down. The function $h_j\equiv h(k_j)$
is defined by
$
h(k)=\frac{k}{2\Gamma}+\frac{\epsilon_d}{2\Gamma}
\left(1-\frac{k}{\omega(k)}\right),
$
where $\Gamma=\pi\rho v^2= v^2/2$ and $\rho=1/2\pi$ is the
density of states of the host metal in the normal state.

In the thermodynamic limit, spin ``rapidities'' $\lambda_\alpha$
are grouped into bound spin complexes of size $n$,
\end{mathletters}
\begin{equation}
\lambda^{(n,j)}_\alpha=\lambda^n_\alpha+i(n+1-2j)/2,\,\,\,
j=1,\ldots,n.
\end{equation}
The simplest bound states of charge excitations are
associated with real spin rapidities $\lambda_\alpha$,
and their charge rapidities $k^{(\pm)}_\alpha$ are
found from the equation
\begin{equation}
h(k_\alpha^{(\pm)})=\lambda_\alpha\pm i/2.
\end{equation}
In the normal state, Eq. (5) has a single solution,
$k_\alpha^{(\pm)}=2\Gamma(\lambda_\alpha\pm i/2)$.
Since $\Delta\ll\Gamma$, this solution acquires in the
superconducting state a small correction of the order of
$(\Delta/\Gamma)^2$,
\begin{mathletters}
\begin{equation}
k^{(\pm)}(\lambda)\simeq 2\Gamma(\lambda\pm i/2)-
\frac{1}{2}\left(\frac{\Delta}{2\Gamma}\right)^2
\frac{\epsilon_d}{(\lambda\pm i/2)^2}.
\end{equation}
The energy of such ``normal'' charge complexes,
$\omega_n(\lambda)=\omega(k^{(+)})+\omega(k^{(-)})$,
is found to be
\begin{equation}
\frac{\omega_n(\lambda)}{4\Gamma}\simeq\lambda+\frac{1}{2}
\left(\frac{\Delta}{2\Gamma}\right)^2
\frac{\lambda-\epsilon_d/2\Gamma}{\lambda^2+1/4},
\end{equation}
where the second term describes the ``gap'' correction. The sign
of this correction is different for different $\lambda$, therefore
the appearance of the energy gap can either increase or decrease
the total energy of normal charge complexes. In what follows, this
fact will play a crucial role in the interplay between the magnetic
and superconducting properties of the system.

Eq. (5) also admits ``gap'' charge complexes with rapidities
$p_\alpha^{(\pm)}=\pm i\Delta \cos{z^{(\pm)}_\alpha}$
and the energy
$\omega_{\rm g}(\lambda_\alpha)=2\Delta{\rm Re}\,\sin{z_\alpha}$.
Here
\end{mathletters}
\begin{equation}
z^{(+)}(\lambda_\alpha)=z^{(-)*}(\lambda_\alpha)=
\arctan{\frac{-i\epsilon_d/2\Gamma}
{\lambda_\alpha-\epsilon_d/2\Gamma+i/2}}
\end{equation}
and the terms of order $\Delta/\Gamma$ are omitted. In what
follows, we consider the case of negative $\epsilon_d$.
The gap complexes can then be shown to exist only for
$\lambda>\epsilon_d/2\Gamma$, and their energies are positive,
$\omega_{\rm g}>0$. BAE admit also ``long'' charge complexes
associated with spin complexes (4). We however do not consider
such excitations, because, as in the normal state, they do not
contribute to the low-temperature thermodynamics of the system.
Finally, the subgap spectrum of the model contains a discrete mode
(DM) \cite{R}, which is naturally treated as a particle-impurity
bound state. In the BA approach, a DM is first found as a solution
of a single particle problem rather than as a multiparticle discrete
mode predicted by Shiba \cite{SH}. It can be shown that for
$\epsilon_d<0$ the renormalized energy of a DM in the multiparticle
spectrum of the model (2) is much bigger than $\Delta$, and hence
a DM does not contribute to thermodynamics of the system in the
temperature range of physical interest, $T\leq T_c$, where $T_c$
is the superconducting critical temperature.

In the standard manner, we then find in the thermodynamic limit
a set of equations for the renormalized energies of the elementary
excitations $\varepsilon(k)$, $\xi(\lambda)$ and $\kappa_n(\lambda)$,
corresponding to unpaired charge excitations with real $k_j$, normal
charge complexes and spin complexes, respectively:
\begin{mathletters}
\begin{eqnarray}
\varepsilon(k)&=&\omega(k)+
\int_{-\infty}^{\infty}d\lambda\,a_1(h(k)-\lambda)\,
F[-\xi(\lambda)]\nonumber\\
&&+\int_{\epsilon_d/2\Gamma}^{\infty}d\lambda\,a_1(h(k)-\lambda)\,
F[-\eta(\lambda)]\nonumber\\
&&-\sum_{n=1}^{\infty}\int_{-\infty}^{\infty}d\lambda\,
a_n(\lambda-h(k))\,F[-\kappa_n(\lambda)]\\
\xi(\lambda)&=&\omega_n(\lambda)
+\int_{-\infty}^{\infty}dk\,h'(k)a_1(\lambda-h(k))F[-\varepsilon(k)]
\nonumber\\
&&+\int_{-\infty}^{\infty}d\lambda a_2(\lambda-\lambda')
F[-\xi(\lambda')]\nonumber\\
&&+\int_{\epsilon_d/2\Gamma}^{\infty}d\lambda\,a_2(\lambda-\lambda')
F[-\eta(\lambda')])\\
F[\kappa_n(\lambda)]&=&
\sum_{m=1}^{\infty}\int_{-\infty}^{\infty}d\lambda\,
A_{nm}(\lambda-\lambda')\,F[-\kappa_m(\lambda')]\nonumber\\
&&+\int_{-\infty}^{\infty}dk\,h'(k)\,
a_n(\lambda-h(k))\,F[-\varepsilon(k)].
\end{eqnarray}
The renormalized energy of gap complexes is given by
\begin{equation}
\eta(\lambda)=\omega_{\rm g}(\lambda)+\xi(\lambda)-\omega_n(\lambda),
\;\;\;\lambda>\epsilon_d/2\Gamma.
\end{equation}
Here, $F[f(x)]\equiv T\ln{[1+\exp{(f(x)/T)}]}$, $h'=dh/dk$,
$a_n(x)=(2n/\pi)(n^2+4x^2)^{-1}$, and $A_{nm}(x)=\delta_{nm}\delta(x)+
(1-\delta_{nm})[a_{|n-m|}+a_{n+m}+
2\sum_{k=1}^{{\rm min}(n,m)-1}a_{|n-m|+2k}(x)]$.
The thermodynamic potentials of the host superconductor,
$\Omega_h$, and the impurity, $\Omega_i$, are found to be
\end{mathletters}
\begin{mathletters}
\begin{eqnarray}
\frac{\Omega_h}{L}&=&\frac{E_{\rm BCS}}{L}
-2\int_{-\infty}^{\infty}\frac{dk}{2\pi}F[-\omega(k)]\\
\Omega_i&=&2\epsilon_d-\xi(\epsilon_d/2\Gamma).
\end{eqnarray}
Therefore, the equation for the order parameter,
$\delta(\Omega_h+\Omega_i)/\delta\Delta=0$, takes the form
\end{mathletters}
\begin{equation}
1=\frac{{\rm g}L}{2}\int_{}^{}\frac{dk}{2\pi}
\frac{\tanh{(\omega(k)/2T)}}{\omega(k)}
-\frac{{\rm g}}{2\Delta}\frac{\delta\Omega_i}{\delta\Delta},
\end{equation}
where the first term is the standard BCS term, while the second
term describes the impurity contribution. The low-temperature
thermodynamics of the $U\to\infty$ Anderson impurity embedded
in a BCS superconductor is thus described completely by
Eqs. (8)-(10).

At finite temperatures, the thermodynamic BA equations require
a numerical analysis. However, at $T=0$ they are significantly
simplified and can be solved analytically. Indeed, one can show
that the energies $\varepsilon(k)$, $\eta(\lambda)$, and
$\kappa_n(\lambda)$ are positive. Therefore, as in the normal
alloys, the ground state of the system contains only normal
charge complexes and is described by a single equation for the
function
$\xi(\lambda)=\xi_-(\lambda)\theta(Q-\lambda)+
\xi_+(\lambda)\theta(\lambda-Q)$, where
$Q$ is a single zero of the function $\xi(\lambda)$,
$\xi(Q)=0$,
\begin{mathletters}
\begin{equation}
\xi(\lambda)=d(\lambda)+
\int_{Q}^{\infty}d\lambda'R(\lambda-\lambda')\xi_+(\lambda').
\end{equation}
Here, $R(x)=\int(d\omega/2\pi)\exp{(-i\omega x)}/(1+\exp{|\omega|})$,
and
\begin{equation}
d(\lambda)=\omega_n(\lambda)-\int_{-\infty}^{\infty}d\lambda'
R(\lambda-\lambda')\omega_n(\lambda').
\end{equation}

Eqs. (11) completely determine the physical properties of
the ground state of the system, but for our purposes it is
more convenient to derive equations describing the ground
state directly from BAE. Introducing the ``particle''
and ``hole'' densities of normal charge complexes,
$\sigma(\lambda)=0$, $\lambda>Q$, and $\tilde{\sigma}(\lambda)=0$,
$\lambda<Q$, respectively, we find in the continuous limit
of BAE \end{mathletters}
\begin{eqnarray}
\frac{1}{2\pi}\frac{dk(\lambda)}{d\lambda}+
\frac{1}{L}a_2(\lambda-\epsilon_d/2\Gamma)&=&
\int_{-\infty}^{Q}d\lambda' a_2(\lambda-\lambda')\sigma(\lambda')
\nonumber\\
&&+\sigma(\lambda)+\tilde{\sigma}(\lambda),
\end{eqnarray}
where $k(\lambda)=k^{(+)}(\lambda)+k^{(-)}(\lambda)$ is the
momentum of complexes. The functions $\sigma(\lambda)$ and
$\tilde{\sigma}(\lambda)$ are divided into the host and impurity
parts, i. e.,
$\sigma(\lambda)=\sigma_h(\lambda)+L^{-1}\sigma_i(\lambda)$,
$\tilde{\sigma}(\lambda)=\tilde{\sigma}_h(\lambda)+
L^{-1}\tilde{\sigma}_i(\lambda)$.
The gap-dependent part of the impurity energy is then given
by
\begin{mathletters}
\begin{equation}
E_i(\Delta)=\frac{\Delta^2}{2\Gamma}\,\Phi,\;\;\;
\Phi=\int_{-\infty}^{Q}d\lambda\phi(\lambda)
\sigma_i(\lambda),
\end{equation}
where $\phi(\lambda)\equiv(\lambda-\epsilon_d/2\Gamma)/(\lambda^2+1/4)$,
while the function $\sigma_i(\lambda)$ is determined by the same
equation as in the normal state,
\begin{equation}
\sigma_i(\lambda)+\tilde{\sigma}_i(\lambda)=
R(\lambda-\epsilon_d/2\Gamma)+
\int_{Q}^{\infty}d\lambda'R(\lambda-\lambda')\tilde{\sigma}_i(\lambda').
\end{equation}

At $T=0$, Eq. (10) reads
\end{mathletters}
\begin{mathletters}
\begin{equation}
1=\frac{{\rm g}L}{2\pi}\int_{0}^{\omega_D}
\frac{dk}{\sqrt{\Delta^2+k^2}}
-\frac{{\rm g}}{2\Delta}\frac{\delta E_i}{\delta\Delta},
\end{equation}
where $\omega_D$ is the Debye frequency. By inserting the
well-known solution of the Wiener-Hopf equation (13b) into
Eq. (13a), we finally obtain
\begin{eqnarray}
E_i(\Delta)&=&-\frac{\Delta^2}{2\Gamma}
\int_{-\infty}^{\infty}\frac{d\omega}{2\pi}
\frac{\phi(-\omega)\exp{(i\omega Q)}}{G^{(-)}(\omega)}
\nonumber\\
&\times&\int_{-\infty}^{\infty}\frac{d\omega'}{2\pi i}
\frac{R(\omega')G^{(-)}(\omega')}{\omega'-\omega+i0}
\exp{(i\omega'\epsilon^*_d/2\Gamma)}.
\end{eqnarray}
Here, $\phi(\omega)$ and $R(\omega)$ are the Fourier images
of the functions $\phi(\lambda)$ and $R(\lambda)$, and
$\epsilon^*_d=\epsilon_d-2\Gamma Q$ is the renormalized
impurity level. The functions $G^{(+)}(\omega)=G^{(-)}(-\omega)$,
$G^{(-)}(2\pi\omega)=\sqrt{2\pi}
[(i\omega+0)/e]^{i\omega}/\Gamma(1/2+i\omega)$ are analytical
functions in the upper $(+)$ and lower $(-)$ half-planes.

Eqs. (14) explicitly determine the order parameter at $T=0$.
Neglecting a small gap correction to the ``normal'' value of
$Q=-(1/2\pi)\ln{({\cal D}^{(+)}/\Gamma)}$, one easily finds
the solution of Eq. (14a) in the multiimpurity case
\end{mathletters}
\begin{equation}
\Delta=\Delta_0\exp{(-\mu_{\rm imp})}.
\end{equation}
The parameter $\mu_{\rm imp}=(c_i\epsilon_F/\Gamma)\,\Phi$,
which is proportional to the impurity concentration $c_i$,
describes the impurity contribution.

The qualitative behavior of the impurity contribution is clearly
seen without a detailed study of the integral in Eq. (14b). Let
us first consider the Kondo limit of the problem, where the
renormalized impurity level lies much below the Fermi energy,
$-\epsilon^*_d/2\Gamma\gg 1$. The function $\phi(\lambda)$
is then positive on an essential interval of integration in
Eq. (13a), $\epsilon_d/2\Gamma<\lambda<Q$, and hence the parameter
$\mu_{\rm imp}$ is also positive. The asymptotic estimate of the
integral (14b) gives $\mu_{\rm imp}\simeq c_i\epsilon_F/|\epsilon_d|$.
Thus, in the Kondo limit, magnetic impurities depress superconductivity.

If the impurity level is shifted to the vicinity of the
mixed-valence regime, $|\epsilon^*_d|/2\Gamma\leq 1$, the
parameter $\mu_{\rm imp}$ changes the sign at some point
$\epsilon_d=\tilde{\epsilon_d}<Q$. At
$\epsilon_d>\tilde{\epsilon_d}$, $\mu_{\rm imp}$ is negative,
and hence the Anderson impurities, which in the mixed-valence
regime play the role of a nonmagnetic resonance energy level
rather than that of a local magnetic moment, enhance
superconductivity.

In summary, making use of the BA approach, we have derived the
exact equations describing the low-temperature thermodynamics
of the model (2). We also have derived an equation for the order
parameter $\Delta$, minimizing the thermodynamic potential of
the system with respect to $\Delta$ \cite{N2}. Finally, at $T=0$
we have evaluated the impurity part of the total energy of the
system, and thus found an exact zero-temperature expression for
the order parameter.

The results obtained have a clear physical meaning. The ground
state of a normal magnetic alloy is well known to be composed
of the charge complexes \cite{BA}. The appearance of a
superconducting energy gap results in a gap correction to the
energy of these complexes, and hence to the impurity part of the
total energy of the system. The sign of the impurity contribution
to an energy balance, $\delta E_i/\delta\Delta$, is different in
the Kondo and mixed-valence regimes, leading respectively to either
a depression or an enhancement of superconductivity. Eq. (15) shows
that at $T=0$ the system remains in the superconducting state at any
impurity concentration. Due to the well-developed Kondo screening
of a local magnetic moment of impurities, the system exhibits a
Cooper pairs weakening rather than a pairs breaking predicted by
the AG theory \cite{AG}, not accounting for the Kondo effect.
Moreover, in the mixed-valence regime, an impurity contribution
to the density of states of the system near the Fermi level even
dominates over a Cooper pairs weakening, and superconductivity is
enhanced.

Finally, it should be emphasized that the results obtained do
not contradict a concept of gapless superconductivity suggested
by Abrikosov and Gor'kov \cite{AG}. At $T=0$ an energy gap in
the spectrum of the system may vanish at some critical impurity
concentration, while an order parameter along with the parameter
$\Delta$ remains finite. The BA technique admits an analytical
computation of an energy gap at $T=0$, that can clarify this
very important and interesting question.

I would like to thank A. A. Abrikosov, S. John, and especially
P. B. Wiegmann for stimulating discussions.


\begin{references}
\bibitem{N}
In a pure BCS superconductor, the parameter $\Delta$ is both an
energy gap and an order parameter. As in Refs. \cite{AG,SMA,NSMA},
we use the term ``order parameter'', because $\Delta$ vanishes in
the normal state of the system. In our case of a strongly correlated
electron system, the parameter $\Delta$ is auxiliary. A real order
parameter of the superconducting transition is only proportional to
$\Delta$, and may differ from both an energy gap in the spectrum of
the system \cite{AG} and from the parameter $\Delta$.
\bibitem{N2}
Eqs. (10) and (14a) are not related to integrability of the model.
They are obviously valid for any dilute magnetic alloy, provided
superconductivity of a host metal is described by the BCS theory.
\end{references}
\end{document}